\documentclass[twocolumn]{aastex63}
\usepackage{graphicx}
\usepackage{color}

\usepackage{amsmath}
\usepackage{amsfonts}
\usepackage{amssymb}
\usepackage{wrapfig}
\usepackage[caption=false]{subfig}
\usepackage{tikz}
\usepackage{enumitem}
\usepackage{relsize}

\graphicspath{{./}{figures/}}

\newcommand{\result}[1]{\textcolor{black}{#1}}

\newcommand{\mmin}{\ensuremath{m_\mathrm{min}}}
\newcommand{\mmax}{\ensuremath{m_\mathrm{max}}}

\newcommand{\alphaonePL}{\result{\ensuremath{-1.52^{+0.45}_{-0.35}}}}

\newcommand{\mminBHPL}{\result{\ensuremath{6.2^{+2.4}_{-4.5} \,M_\odot}}}
\newcommand{\alphaBHPL}{\result{\ensuremath{-1.34^{+0.87}_{-0.80}}}}
\newcommand{\mmaxBHPL}{\result{\ensuremath{42.2^{+20.2}_{-5.5} \,M_\odot}}}
\newcommand{\betaBHPL}{\result{\ensuremath{7.2^{+4.4}_{-5.4}}}}

\newcommand{\SDDRgap}{\result{4.6}}
\newcommand{\Palphagtralpha}{\result{91\%}}

\newcommand{\LVCBNSrate}{\result{\ensuremath{{1540}^{+3200}_{-1220}}}}
\newcommand{\LVCBBHrate}{\result{\ensuremath{{53.2}^{+55.8}_{-28.2}}}}

\newcommand{\BNSratePL}{\result{\ensuremath{199^{+817}_{-173}}}}

\newcommand{\BNSratedipbreak}{\result{\ensuremath{871^{+3015}_{-805}}}}

\newcommand{\BBHratePL}{\result{\ensuremath{77.7^{+60.6}_{-38.5}}}}

\newcommand{\BBHratedipbreak}{\result{\ensuremath{47.5^{+57.9}_{-28.8}}}}

\newcommand{\rateunits}{\ensuremath{\ \mathrm{Gpc}^{-3} \ \mathrm{yr}^{-1}}}

\begin{document}
%-------------------------------------------------

\title{Does Matter Matter? Using the mass distribution to distinguish neutron stars and black holes}

\author{Maya Fishbach}
\affiliation{Department of Astronomy and Astrophysics, University of Chicago, Chicago, IL 60637, USA}
\author{Reed Essick}
\affiliation{Kavli Institute for Cosmological Physics, University of Chicago, Chicago, IL 60637, USA}
\author{Daniel E. Holz}
\affiliation{Department of Astronomy and Astrophysics, University of Chicago, Chicago, IL 60637, USA}
\affiliation{Kavli Institute for Cosmological Physics, University of Chicago, Chicago, IL 60637, USA}
\affiliation{Enrico Fermi Institute, University of Chicago, Chicago, IL 60637, USA}
\affiliation{Department of Physics, University of Chicago, Chicago, IL 60637, USA}

\begin{abstract}
Gravitational-wave detectors have opened a new window through which we can observe black holes (BHs) and neutron stars (NSs).
Analyzing the 11 detections from LIGO/Virgo's first gravitational-wave catalog, GWTC-1, we investigate whether the power-law fit to the BH mass spectrum can also accommodate the binary neutron star (BNS) event GW170817, or whether we require an additional feature, such as a mass gap, in between the NS and BH populations.
We find that with respect to the power-law fit to binary black hole (BBH) masses, GW170817 is an outlier at the \result{0.13\%} level, suggesting a distinction between NS and BH masses.
A single power-law fit across the entire mass range is in mild tension with: (a) the detection of one source in the BNS mass range ($\sim 1$--$2.5 \,M_\odot$), (b) the absence of detections in the ``mass-gap'' range ($\sim 2.5$--$5 \,M_\odot$), and (c) the detection of 10 sources in the BBH mass range ($\gtrsim 5 \,M_\odot$).
Instead, the data favor models with a feature between NS and BH masses, including a mass gap (Bayes factor of \SDDRgap{})
and a break in the power law, with a steeper slope at NS masses compared to BH masses (\result{91\%} credibility).
We estimate the merger rates of compact binaries based on our fit to the global mass distribution, finding $\mathcal{R}_\mathrm{BNS} = \BNSratedipbreak \rateunits$ and $\mathcal{R}_\mathrm{BBH} = \BBHratedipbreak \rateunits$.
We conclude that, even in the absence of any prior knowledge of the difference between NSs and BHs, the gravitational-wave data alone already suggest two distinct populations of compact objects.
\end{abstract}

%-------------------------------------------------
\section{Introduction}
\label{sec:introduction}

The mass distribution of neutron stars (NSs) and stellar-mass black holes (BHs) is fundamental to our understanding of stellar evolution, {binary formation channels}, supernova physics, and the nuclear equation of state (EoS).
There has been considerable effort to measure the mass distribution for NSs and BHs based on radio, X-ray, and optical observations of these systems~\citep{2011MNRAS.414.1427V,2012ApJ...757...55O,2013ApJ...778...66K,2016arXiv160501665A,2018MNRAS.478.1377A,2019ApJ...876...18F,2020RNAAS...4...65F}.
Indeed, there are several features in the mass distribution that are particularly relevant for understanding the physics of these systems, including the maximum NS mass, the minimum BH mass, and the purported mass gap between the most massive NS and the least massive BH.
The maximum possible NS mass is governed by the nuclear EoS, and there has been significant work to extract this value by measuring the masses of electromagnetically identified NSs (see \citealt{2012ARNPS..62..485L} for a review).
The maximum mass of the astrophysical population of NSs is currently estimated to be $\sim 2$--$2.6 \,M_\odot$~\citep{2016arXiv160501665A, 2018MNRAS.478.1377A, 2020RNAAS...4...65F}.
%Although the maximum mass among astrophysically occurring NSs in binary systems may, in general, differ from the maximum gravitational mass supported by the nuclear EoS \citep[see, e.g., discussions in ][]{Miller_2019,2020arXiv200304880L}, it provides a useful lower bound on this uncertain quantity.
Meanwhile, analyses of the BH mass distribution based on the sample of $\sim 20$ BHs in X-ray binary systems suggest that the minimum BH mass does not coincide with the maximum NS mass, implying that there is a mass gap between the two populations \citep{2010ApJ...725.1918O,2011ApJ...741..103F}.
However, it has been proposed that this observed mass gap may not be physical, but rather an artifact of X-ray selection effects \citep{2012ApJ...757...36K}. Recently, a low-mass BH, possibly occupying the mass gap, was discovered in radial velocity searches~\citep{2019Sci...366..637T}, \added{and a candidate mass-gap BH was discovered in the compact binary system GW190814~\citep{GW190814}\footnote{The secondary component of GW190814, with mass $m_2 = 2.59^{+0.08}_{-0.09} \ M_\odot$ may alternatively be the most massive NS ever observed.}.}
Understanding whether or not there is a mass gap between NSs and BHs in binary systems has implications for supernova theory and binary physics~\citep{2001ApJ...554..548F,2012ApJ...757...91B, 2019ApJ...878L...4B}.

Gravitational-wave (GW) detections by Advanced LIGO~\citep{2015CQGra..32g4001L} and Virgo~\citep{2015CQGra..32b4001A} provide a rapidly growing sample of binary black hole (BBH) and binary neutron star (BNS) systems.
Analyzing the masses of these detections can provide a measurement of the maximum NS mass \citep{2020arXiv200500482C} and identify the presence of a mass gap between neutron stars and black holes \citep{2015ApJ...807L..24L,2015MNRAS.450L..85M,2017MNRAS.465.3254M, 2017PhRvD..95j3010K}.
This measurement is challenging because large observational uncertainties for the component masses often make it difficult to determine whether individual systems are in the NS mass range, the mass gap, or the BH mass range~\citep{2013ApJ...766L..14H,2015ApJ...807L..24L, 2015MNRAS.450L..85M}.
\citet{2015ApJ...807L..24L} and \citet{2015MNRAS.450L..85M,2017MNRAS.465.3254M} found that $\sim 100$ low-mass detections are required to confidently detect the presence of a mass gap and measure the maximum NS mass and minimum BH mass if these features are sharp.
Alternatively, it has been proposed that tidal information encoded in the GW signal can be used to distinguish populations of BBH, BNS, and neutron star- black hole (NSBH) systems~\citep{PhysRevD.77.021502,2013PhRvD..88d4042R,2020ApJ...893L..41C,2020arXiv200501726F}, and \cite{2020arXiv200101747W} recently proposed an analysis to jointly measure the tidal deformability and derived quantities like the EoS together with the mass and spin distribution of the BNS population.
However, the imprint of tides is much harder to extract from the GW signal than the masses~\citep{2015PhRvD..91d3002L}.

In this paper, we focus on the mass distribution alone and characterize a possible mass feature, such as a gap, between the BNS and BBH populations.
To do this, we jointly analyze the masses of the 10 BBH systems and one BNS system detected by the LIGO/Virgo Collaboration (LVC) in their first two observing runs (O1 and O2) and published in the catalog GWTC-1~\citep{2019PhRvX...9c1040A}.
We thereby explore whether GW170817, the one BNS system of GWTC-1, is distinguishable from the BBH population based only on its mass.

In addition to the events published by the LVC in GWTC-1, new candidate BBHs have been identified in the public O1 and O2 data~\citep{2019ApJ...872..195N, 2020PhRvD.101h3030V, 2020ApJ...891..123N}.
In order to ensure that we understand the selection function for the catalog (see \S\ref{sec:statmethods}), we do not analyze these additional systems here, but given that they are relatively high-mass BBHs, we would not expect their inclusion to change our main conclusions.

\added{Furthermore, three events from the third observing run (O3) have been published by the LVC to date: GW190425, GW190412, and GW190814~\citep{GW190425, GW190412, GW190814}. Both GW190425, a system with a total mass of $\sim 3 \ M_\odot$ and GW190814, a system with a secondary mass of $\sim 2.6 \ M_\odot$, are directly pertinent to the subject of this work, as they feature systems that may fall within the mass gap.
Without the context of the full set of O3 events, we cannot yet include these additional systems in our population analysis.
However, it is clear that the methods described here will be relevant when analyzing events from O3 and beyond.}

The remainder of the paper is organized as follows.
\S\ref{sec:methods} describes the technical details of the analysis, including the parameterization employed for the mass distribution (\S\ref{sec:models}) and the statistical framework of the population analysis (\S\ref{sec:statmethods}).
\S\ref{sec:powerlawoutlier} explores the extension of a BBH power-law fit down to the BNS mass range.
We find that a single power law struggles to simultaneously fit the relatively high rate of detections in the BNS mass range (one) compared to BBH detections (10) and the lack of detections in between.
In \S\ref{sec:results} we fit for {possible} features between the NS and BH mass range, including a dip and/or break in the power law, and quantify the preference for these features.
In \S\ref{sec:discussion} we discuss how our results can be used to classify detections into NS and BH categories (\S\ref{sec:single-event}) and to infer the merger rate of BNS, NSBH, and BBH systems (\S\ref{sec:rates}), as well as future prospects (\S\ref{sec:future}).
We conclude in \S\ref{sec:conclusion}.

%-------------------------------------------------
\section{Methods}
\label{sec:methods}

We describe the parameterization of the mass distribution in \S\ref{sec:models} and then discuss the statistical framework upon which we base our inference in \S\ref{sec:statmethods}.

%------------------------
\subsection{Mass model}
\label{sec:models}

For our simplest model of the component mass distribution, we consider a power law with a variable minimum mass $\mmin$, slope $\alpha$, and maximum mass $\mmax$ \citep{2017ApJ...851L..25F,2017PhRvD..95j3010K,2019ApJ...882L..24A,2019PhRvD.100d3012W}:
\begin{equation}\label{eq:pm-pl}
    p_\mathrm{PL}(m) \propto
    \begin{cases}
         m^\alpha & \text{if } \mmin < m < \mmax \\
         0 & \text{else.}
    \end{cases}
\end{equation}
This \textsc{power-law} adequately fits the BH mass distribution as inferred from the GWTC-1 BBH detections~\citep{2019ApJ...882L..24A, 2020ApJ...891L..31F}.
When adding the BNS detection to the fit, we gradually build on top of this simple mass distribution, introducing phenomenological features to capture possible deviations from a pure power-law.

\begin{figure}
    \begin{center}
        \includegraphics[width=0.9\columnwidth, trim=100 265 505 125, clip]{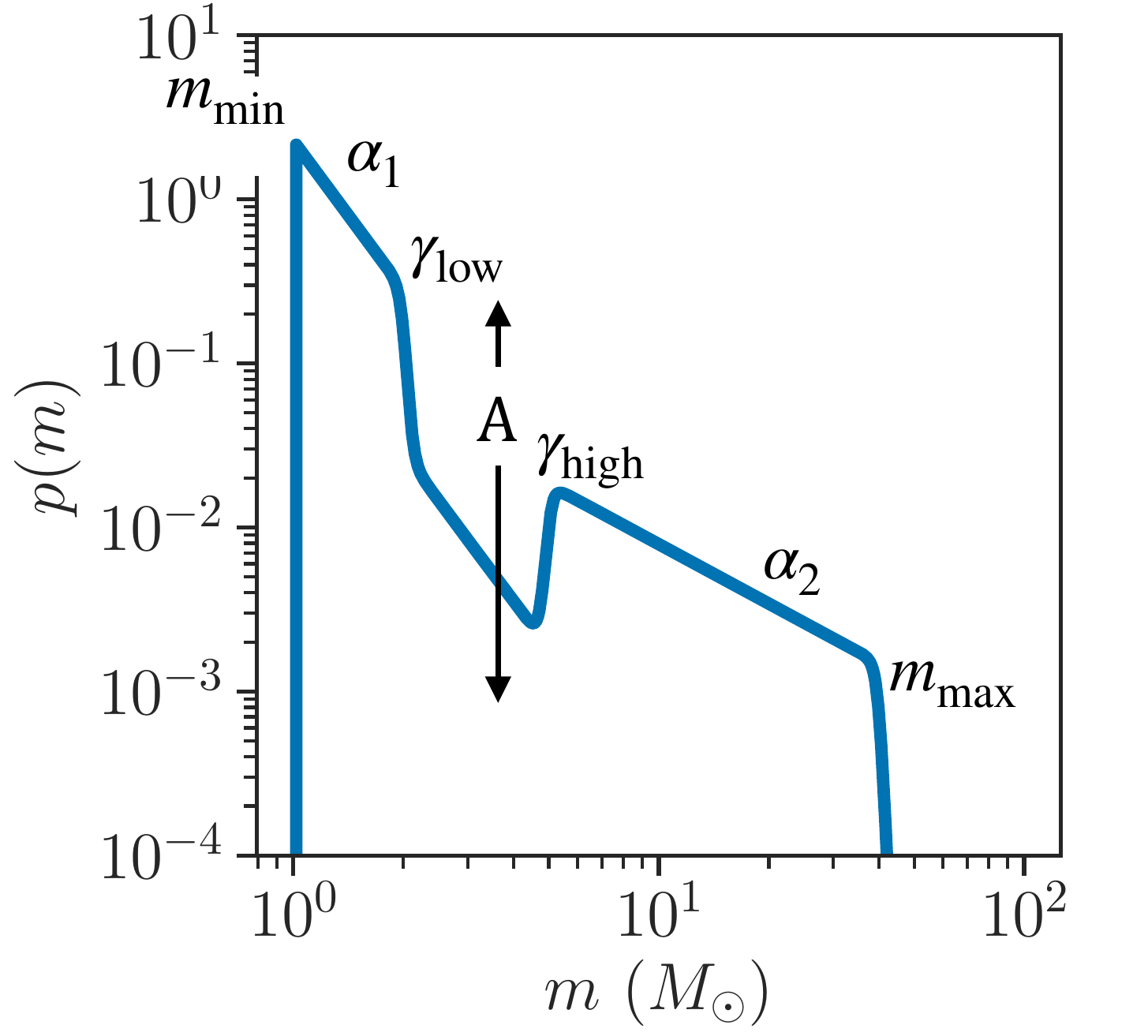}
        \includegraphics[width=0.9\columnwidth, trim=505 265 100 125, clip]{massdist_example_1d-2d-annotated.pdf}
    \end{center}
    \caption{
        Example phenomenological distribution described in \S\ref{sec:models}.
        \emph{Top}: the one-dimensional mass distribution parametrized according to Eq.~\ref{eq:pm-dipbreak}: a broken power-law with slopes $\alpha_1$ and $\alpha_2$ and break at $\gamma_\mathrm{high}$, with a notch filter between $\gamma_\mathrm{low}$ and $\gamma_\mathrm{high}$ with amplitude $A$.
        \emph{Bottom}: the corresponding two-dimensional distribution, constructed from the one-dimensional distribution with a mass-ratio dependent pairing function following Eq.~\ref{eq:pm-2d}. The colorbar denotes the probability density $p(m_1, m_2)$.
    }
    \label{fig:example distribution}
\end{figure}

To allow for the possibility of a dip or gap in the mass spectrum, we multiply the original \textsc{power-law} mass spectrum by a notch filter,
\added{
\begin{equation}\label{eq:notch}
    n(m) = 1 - \frac{A} {\left(1 + \left(\frac{\gamma_\mathrm{low}}{m}\right)^{\eta_\mathrm{low}}\right) \left(1 + \left(\frac{m}{\gamma_\mathrm{high}}\right)^{\eta_\mathrm{high}}\right)},
\end{equation}
}
which suppresses the distribution when $\gamma_\mathrm{low} < m < \gamma_\mathrm{high}$.
We refer to this model as \textsc{power-law + dip}.
The parameters $\eta_\mathrm{low}$ and $\eta_\mathrm{high}$ set the sharpness of the dip's edges, while the amplitude of the dip is set by the parameter $A$.
In principle we can allow the data to inform our knowledge of the sharpness of the gap edges in addition to their placement and the depth of the gap.
However, since we cannot meaningfully constrain all of these features with only 11 events, we fix the edges to be near-infinitely sharp: $\eta_\mathrm{low} = \eta_\mathrm{high} = 50$.
With sharp edges, $A = 1$ corresponds to an empty gap, while $A = 0$ corresponds to no dip. $A < 0$ corresponds to a bump rather than a dip. \added{We verify that the mass distribution we infer by simultaneously fitting the sharpness parameters $\eta_\mathrm{low}$ and $\eta_\mathrm{high}$, in addition to $\gamma_\mathrm{low}$, $\gamma_\mathrm{high}$, and $A$, is nearly identical to the mass distribution inferred under the reduced model with $\eta_\mathrm{low} = \eta_\mathrm{high} = 50$. This is due to the degeneracies between the parameters of Eq.~\ref{eq:notch} within the large statistical uncertainties on the shape of the gap.}

Similar to the notch filter that models the mass gap, we can use a low-pass filter that ``turns off" the mass distribution when $\mmax < m$ to model the edge of the upper/pair-instability mass gap, rather than the sharp cutoff at $\mmax$.
In other words, we can replace the condition that $p(m > \mmax) = 0$ by multiplying $p(m)$ by:
\begin{equation}
    l(m) = \left(1 + \left(\frac{m}{\mmax}\right)^n \right)^{-1},
\end{equation}
where large $n$ corresponds to a sharp cutoff and small $n$ corresponds to a gradual turn-off.
Similar to the low-mass gap's edges, we do not have enough detections to meaningfully constrain the steepness of the upper mass gap, and we fix the cutoff to be sharp: $n = 50$.

As a final complication, we include the possibility of a break in the power-law at $\gamma_\mathrm{high}$, so that objects below the gap may follow a different power-law slope $\alpha_1$ from objects above the gap with slope $\alpha_2$:
\begin{equation}
    p_\mathrm{BPL}(m) \propto
    \begin{cases}
        m^{\alpha_1} & \text{if } \mmin < m \leq \gamma_\mathrm{high} \\
        m^{\alpha_2} & \text{if } \gamma_\mathrm{high} < m < \mmax \\
        0 & \text{else.}
    \end{cases}
\end{equation}
The most general one-dimensional mass distribution we consider is therefore:
\begin{equation}\label{eq:pm-dipbreak}
    p(m \mid \lambda) \propto p_\mathrm{BPL}(m) \times n(m) \times l(m),
\end{equation}
with free parameters $\{\mmin, \mmax, \alpha_1, \alpha_2, A, \gamma_\mathrm{low}, \gamma_\mathrm{high}\}$ denoted by $\lambda$.
We refer to this as \textsc{power-law + dip + break}.

Figure~\ref{fig:example distribution} demonstrates some of the features of our parameterization.
Physically, $\gamma_\mathrm{low}$ may correspond to the maximum NS mass and $\gamma_\mathrm{high}$ to the minimum BH mass.
However, the mapping between the physical properties, such as the maximum NS mass, and features in the overall mass distribution, such as the onset of a mass gap, may be more complicated due to, for example, the supernova mechanism or accretion from a binary partner.
This idealized model allows us to explore whether a single power-law ($A = 0$; $\alpha_1 = \alpha_2$) can fit the entire compact object mass spectrum, or whether there is a distinguishing feature between the NS and BH mass spectrum in the form of a dip ($0 < A < 1$), a gap ($A = 1$), and/or a break in the power-law ($\alpha_1 \neq \alpha_2$).
If such a feature is found, it can be used to identify sub-populations of the compact object mass distribution.
In this case, the total mass distribution can alternatively be modeled as a mixture model of sub-populations (e.g. \citealt{2020CQGra..37d5007K}).

As in \citet{2020ApJ...891L..27F} and \citet{2020ApJ...893...35D}, we assume a simple pairing function to generate a joint distribution over both component masses that make up a binary system, given a particular component mass distribution:
\begin{align}
    p(m_1, & m_2 \mid \Lambda) \propto \nonumber \\
        & p(m = m_1 \mid \lambda)\, p(m = m_2 \mid \lambda) \left(\frac{m_2}{m_1}\right)^\beta \Theta(m_2 \leq m_1), \label{eq:pm-2d}
\end{align}
where $\Lambda$ represents the total set of free parameters $\{\mmin, \mmax, \alpha_1, \alpha_2, A, \gamma_\mathrm{low}, \gamma_\mathrm{high}, \beta \}$, or the union of $\lambda$ and $\{\beta\}$, and $\Theta$ is the Heaviside step function that enforces our labeling convention that $m_2 \leq m_1$.
Here we take the pairing function to be a power-law in the mass ratio.
More complicated pairing probabilities are possible, and indeed, any function $p(m_1, m_2)$ can be factored into a product of the one-dimensional mass distribution and a pairing function. We stick with this simple model because it adequately reproduces the observed distribution of GWTC-1~\citep{2020ApJ...891L..27F}.

%------------------------
\subsection{Statistical framework}
\label{sec:statmethods}

The analysis presented in this work consists of two main steps: (a) \emph{model fitting,} i.e., given the GW data, measuring the population parameters of the mass distribution model; and (b) \emph{model checking}, i.e., simulating sets of observable data from the fit to the model and evaluating how closely they resemble the actual set of observed data. This subsection provides an overview of these calculations; the details are provided in the Appendix.

Using the parametrized mass distributions from \S\ref{sec:models}, $p(m_1, m_2 \mid \Lambda)$, we construct a hierarchical Bayesian inference to determine the appropriate population-level parameters, $\Lambda$, given the observed data $\{D_i\}$~\citep{2004AIPC..735..195L,2010PhRvD..81h4029M,2019MNRAS.486.1086M,2019PASA...36...10T}. We focus on the mass distribution alone, fixing the spin distribution (uniform in spin magnitude and isotropic in orientation) and the redshift distribution (flat in comoving volume and source-frame time).
The posterior on the population hyper-parameters, $p(\Lambda \mid \{D_i\})$, is evaluated according to the methods in Appendix~\ref{sec:statmethods-appendix}.
Each draw from this hyper-posterior $p(\Lambda \mid \{D_i\})$ corresponds to a mass distribution $p(m_1, m_2 \mid \Lambda)$.
Averaging the mass distribution over the hyper-posterior yields the posterior population distribution:
\begin{equation}
p(m_1, m_2 \mid \{D_i\}) = \int p(m_1, m_2 \mid \Lambda) p(\Lambda \mid \{D_i\}) d\Lambda.
\end{equation}

We often present the mass distribution in terms of the astrophysical merger rate density, denoted by ${d{N}}/{dm_1 dm_2 dV_c dt_s}$, where $V_c$ is the comoving volume\footnote{We adopt the {\em Planck}\/ 2015 cosmology throughout~\citep{Planck:2015,2018AJ....156..123A}.} and $t_s$ is the time as measured in the source-frame, rather than the probability density $p(m_1, m_2)$.
While the probability density $p(m_1, m_2)$ integrates to unity, the rate density integrated over the masses $(m_1, m_2)$ yields the overall merger rate, ${d{N}}/{dV_c dt_s}$.
The probability density and the rate density are related according to Eq.~\ref{eq:numberdensity}--\ref{eq:NtoR}.

Once we fit the population model, we perform a posterior predictive check by comparing the distribution of \emph{observed} masses as implied by the fit to the model, or the posterior predictive distribution (PPD), to the actual set of observed events, or the empirical distribution function (EDF). This comparison provides strong goodness-of-fit tests, along with providing further insight into why certain features of the overall mass distribution are favored.

%-------------------------------------------------
\section{Can a single power law fit NS and BH masses?}
\label{sec:powerlawoutlier}

In this section we discuss the ability of a simple power law (\textsc{power-law} of Eq.~\ref{eq:pm-pl}) to reproduce the 11 detections of GWTC-1.
We ask whether the BNS detection, GW170817, is distinguishable from the BBH population based on its mass alone. \
If we did not know (based, for example, on its electromagnetic counterpart or prior astrophysical information) that GW170817 belonged to a separate class of compact objects, would we have classified it as a population outlier based on its mass?
{\it Do the gravitational-wave data, alone, suggest the existence of distinct populations of neutron stars and black holes?}

\begin{figure*}
    \begin{center}
        \includegraphics[width=0.90\textwidth]{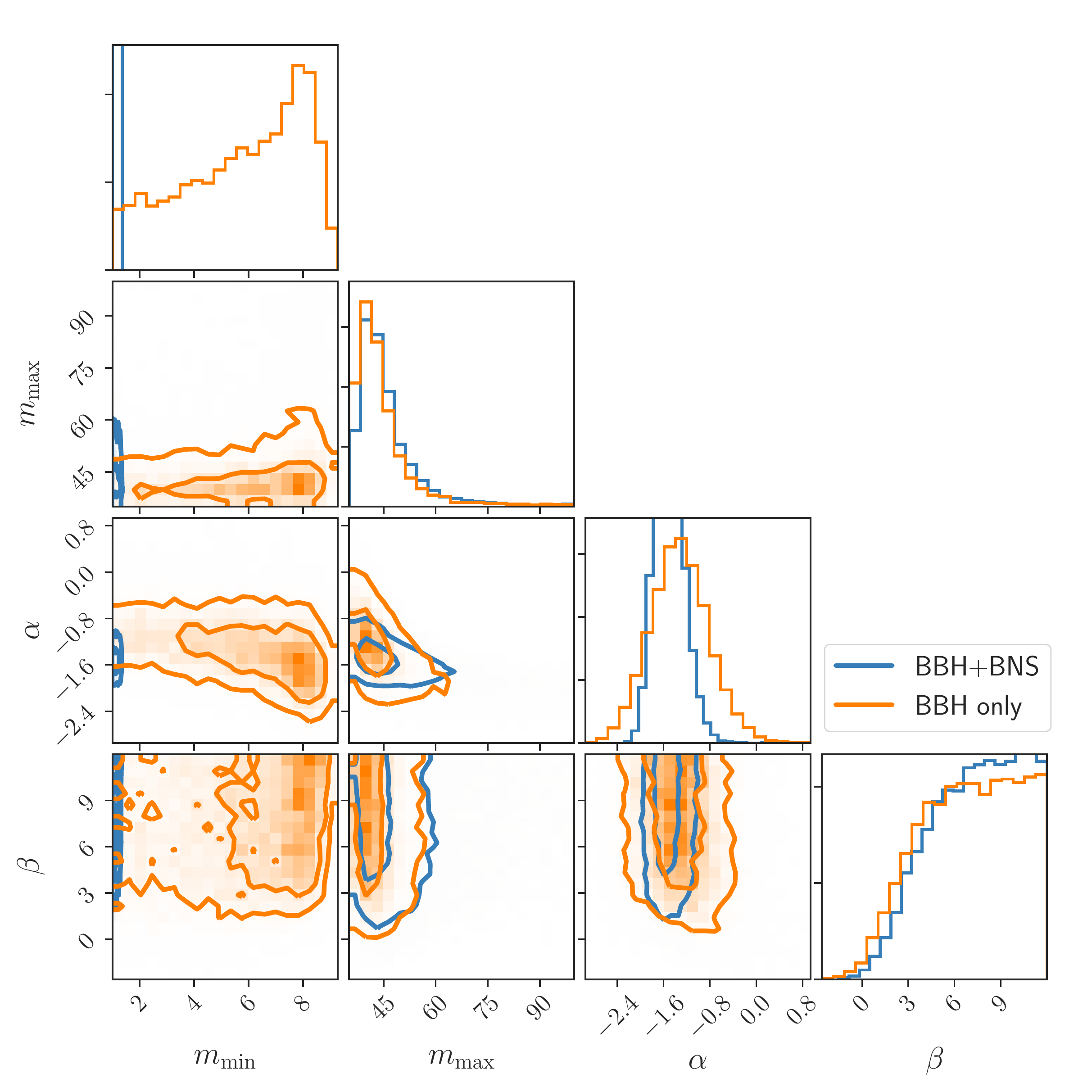}
    \end{center}
    \caption{
        Corner plot~\citep{corner} comparing (\emph{orange}) the power-law fit to the 10 BBH and (\emph{blue}) the fit to all 11 events.
        Contours show 50\% and 90\% credible regions.
        The main effect of adding the BNS event GW170817 to the power-law fit is on the $\mmin$ constraints ({\emph{first column}}), particularly the joint $\alpha$--$\mmin$ constraints ({\emph{first column, third row}}).
        The low mass of GW170817 forces $\mmin \lesssim 1.3$, but prefers a relatively steeper power-law slope compared to the BBH-only joint fit for $\alpha$--$\mmin$.
        However, the power-law fit to all 11 events remains consistent with the BBH-only fit within the 90\% level.}
    \label{fig:PL_corner_compare}
\end{figure*}

We begin by exploring whether the same power-law that fits the BBH detections can also accommodate the BNS detection, GW170817.
The \textsc{power-law} fit to the 10 GWTC-1 BBH yields $\mmin = \mminBHPL$, $\alpha = \alphaBHPL$, $\mmax = \mmaxBHPL$, and $\beta  = \betaBHPL$ (median and 90\% equal-tailed intervals; see Fig.~\ref{fig:PL_corner_compare}).
Unlike previous power-law fits to the BBH~\citep{2019ApJ...882L..24A,2019MNRAS.484.4216R,2020ApJ...891L..27F}, we allow the prior on the minimum mass to extend down to $1\,M_\odot$, as opposed to $3\,M_\odot$ or $5\,M_\odot$.
Specifically, we assume flat priors on all hyper-parameters in the range \result{$\mmin \in  [1, 10]$, $\mmax \in [30, 100]$, $\alpha \in [-4,2]$, $\beta \in [-4, 12]$}.
We find that although the posterior on $\mmin$ peaks at $8.2 \,M_\odot$, there remains posterior support down to the lower prior boundary of $\mmin = 1 \,M_\odot$.
The posterior probability at the peak \result{$\mmin = 8.2 \,M_\odot$} is \result{$\sim 3$} times larger than at \result{$\mmin = 1 \,M_\odot$}.

While the \textsc{power-law} fit to the BBH does not rule out masses as low as $1 \,M_\odot$, we would not expect to detect them very often. Based on the BBH-only \textsc{power-law} fit, we would expect to detect a GW170817-like system, with primary mass $m_1 \lesssim 2 \,M_\odot$, in a set of 11 total detections only \result{0.13\%} of the time, suggesting that GW170817 is a fairly atypical system with respect to the BBH population.

If we now include GW170817 and fit the \textsc{power-law} model to all 11 events in GWTC-1, we find constraints on the hyper-parameters that are broadly compatible with the BBH-only fit; see the comparison in Fig.~\ref{fig:PL_corner_compare}.
The largest shift between the BBH-only fit and the all-event fit is in the joint ($\mmin$, $\alpha$) posterior, as seen in the third row, first column of Fig.~\ref{fig:PL_corner_compare}.
Due to the correlation between $\alpha$ and $\mmin$, if we constrain $\mmin < 2 \,M_\odot$ in the BBH-only fit, we find a shallower slope, \result{$\alpha = -1.09^{+0.79}_{-0.56}$}, compared to the slope in the all-events fit, $\alpha = \alphaonePL$.
Nevertheless, the hyper-parameter posteriors agree within the 90\% levels between the two fits.

\begin{figure}[t]
    \begin{center}
        \includegraphics[width=1.0\columnwidth, clip=True, trim=0 20 22 20]{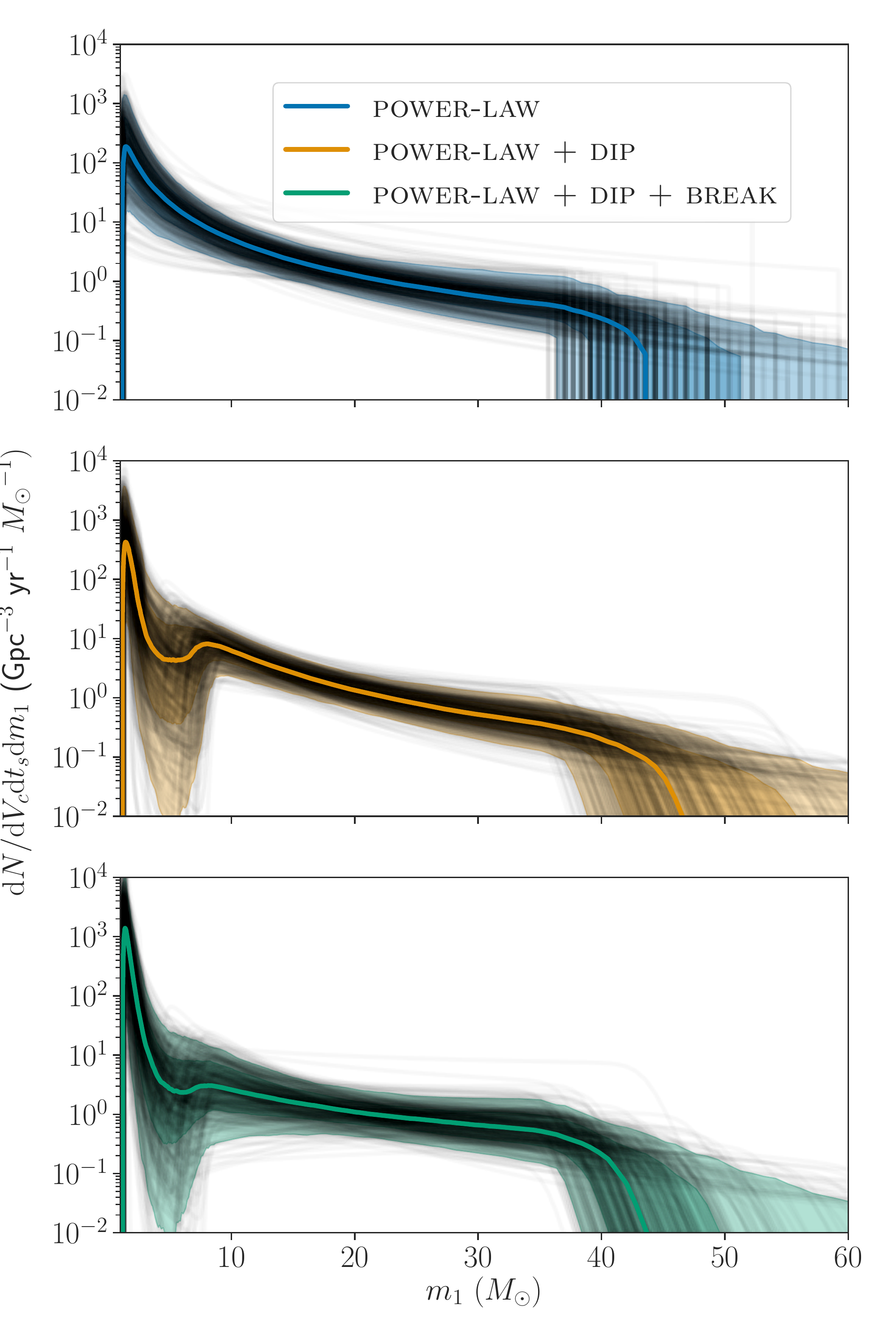}
    \end{center}
    \caption{
        Posterior population distributions for three models, in order of increasing complexity: \textsc{power-law}, \textsc{power-law + dip}, \textsc{power-law + dip + break}.
        Each panel shows the differential merger rate density, $dN/dV_c dt_s dm_1$, as a function of primary mass $m_1$.
        The colored lines show the median rate density, and the colored shaded bands show 1-$\sigma$ (68\%) and 2-$\sigma$ (95\%) credible intervals.
        In gray, we plot 500 draws from the population posterior under each model.
    }
    \label{fig:dragons}
\end{figure}

\begin{figure}[h!]
\captionsetup[subfloat]{captionskip=0cm}
    \begin{center}
        \begin{tikzpicture}
            \draw (+0, +0) node (ppc-pl) {\includegraphics[width=0.99\columnwidth, trim=3 45 5 8, clip]{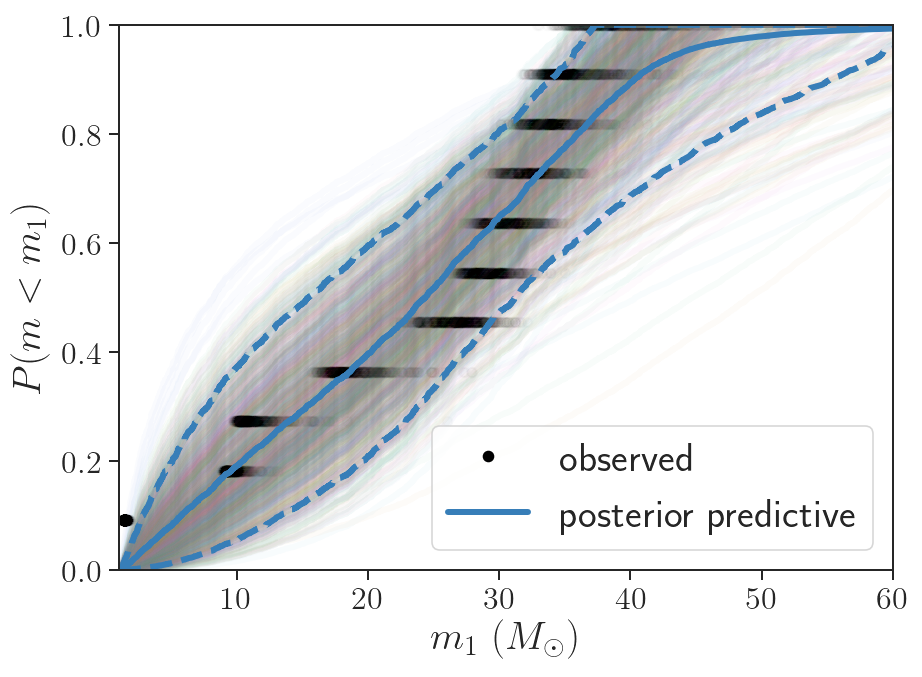}};
            \draw (-1.9, +2.3) node {(a) \textsc{power-law}};
            \draw (+0, -5.4) node (ppc-dip) {\includegraphics[width=0.99\columnwidth, trim=3 45 5 8, clip]{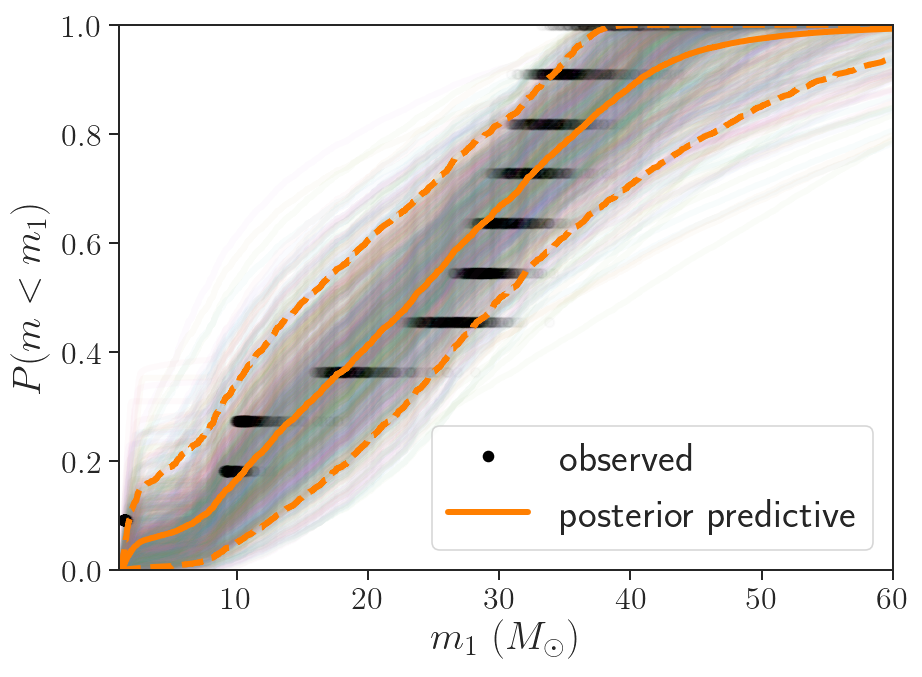}};
            \draw (-1.4, -3.1) node {(b) \textsc{power-law + dip}};
            \draw (+0, -11.1) node (ppc-breakdip) {\includegraphics[width=0.99\columnwidth, trim=3 10 5 8, clip]{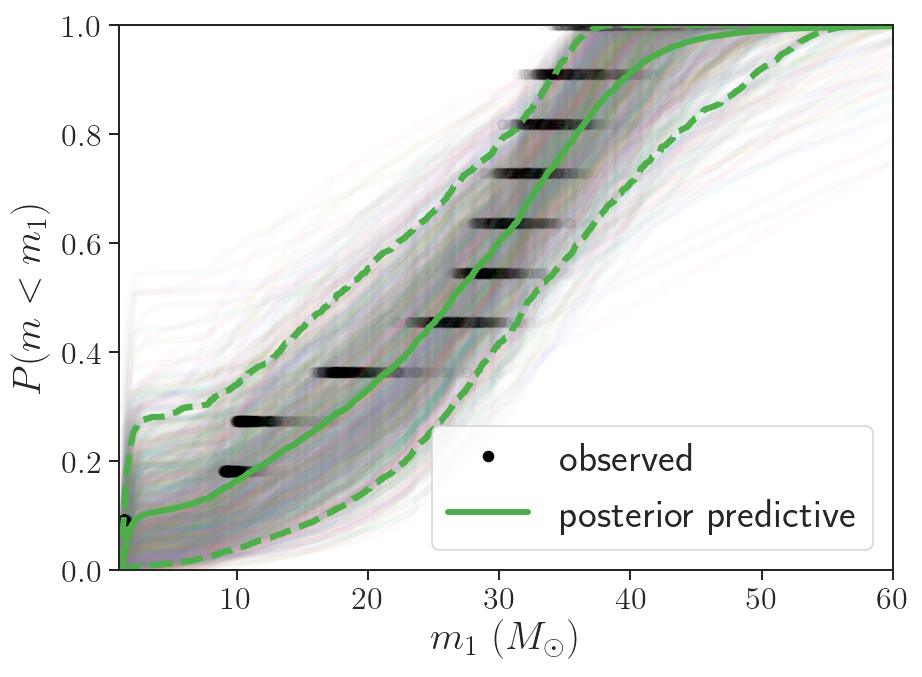}};
            \draw (-0.7, -8.50) node {(c) \textsc{power-law + dip + break}};
        \end{tikzpicture}
    \end{center}
    \vspace{-0.75cm}
    \caption{
        Posterior predictive check comparing the observed primary mass distribution as predicted from our models (\emph{thin colored curves}; each curve corresponds to a draw from the hyper-parameter posterior) with the empirical distribution from the 11 GWTC-1 events (\emph{black}; each point corresponds to a draw from the population-informed single-event posterior).
        The solid colored line in each panel corresponds to the posterior average (mean) of the predicted distributions, while the two dashed colored lines denote the symmetric 90\% interval around the predicted curves.
        The model is a good fit to the data if the empirical distribution (\emph{black}) is contained within the range of model predictions (\emph{colored}). The top panel shows that the \textsc{power-law} model has trouble accounting for GW170817, while the other models account for all 11 events.
    }
    \label{fig:ppc}
\end{figure}
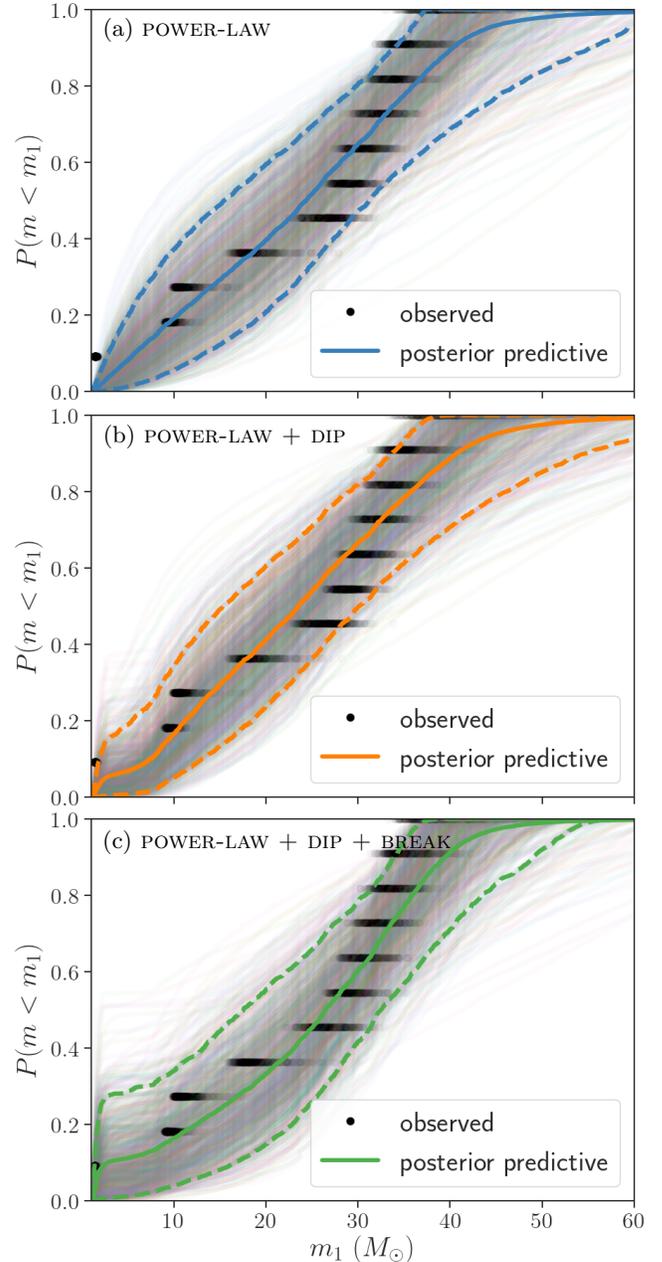

The inferred primary mass distribution for the \textsc{power-law} fit to the 11 GWTC-1 events can be seen in the top panel of Fig.~\ref{fig:dragons}, while the goodness of fit of the \textsc{power-law} model can be visualized in the posterior predictive check of Fig.~\ref{fig:ppc}.
The thin colored curves of Fig.~\ref{fig:ppc} show 1,000 draws from the PPD of \emph{observed} primary masses.
Each curve corresponds to a different draw from the hyper-parameter posterior.
The solid line corresponds to the mean of these 1,000 realizations, and the dashed lines show 90\% credible bounds.
In black, we show 1,000 draws from the empirical cumulative distribution, or empirical distribution function (EDF).
Each draw from the EDF is found by {reweighting} the single-event posteriors to the population prior.
From the updated posterior for each of the 11 events, we draw an $m_1$ sample, and order these 11 points from smallest to largest.
The EDF passes through \result{$(\sim 1.6 \,M_\odot, 1/11)$}, driven by the primary mass of GW170817, which is above the {90\%} predictive band.
While this is suggestive, it is not terribly unexpected from noise fluctuations affecting the most extreme members of a set~\citep{1951JAM....18..293W, 2020ApJ...891L..31F}.
Based on the \textsc{power-law} fit to all events, we expect to detect an event with primary mass $m_1 < 2 \,M_\odot$ in a set of 11 events \result{17\%} of the time, {a significant} shift from the expected \result{0.13\%} for the BBH-only fit.

\begin{figure*}
    \begin{center}
        \includegraphics[width=\textwidth, clip=True, trim=0 10 0 0]{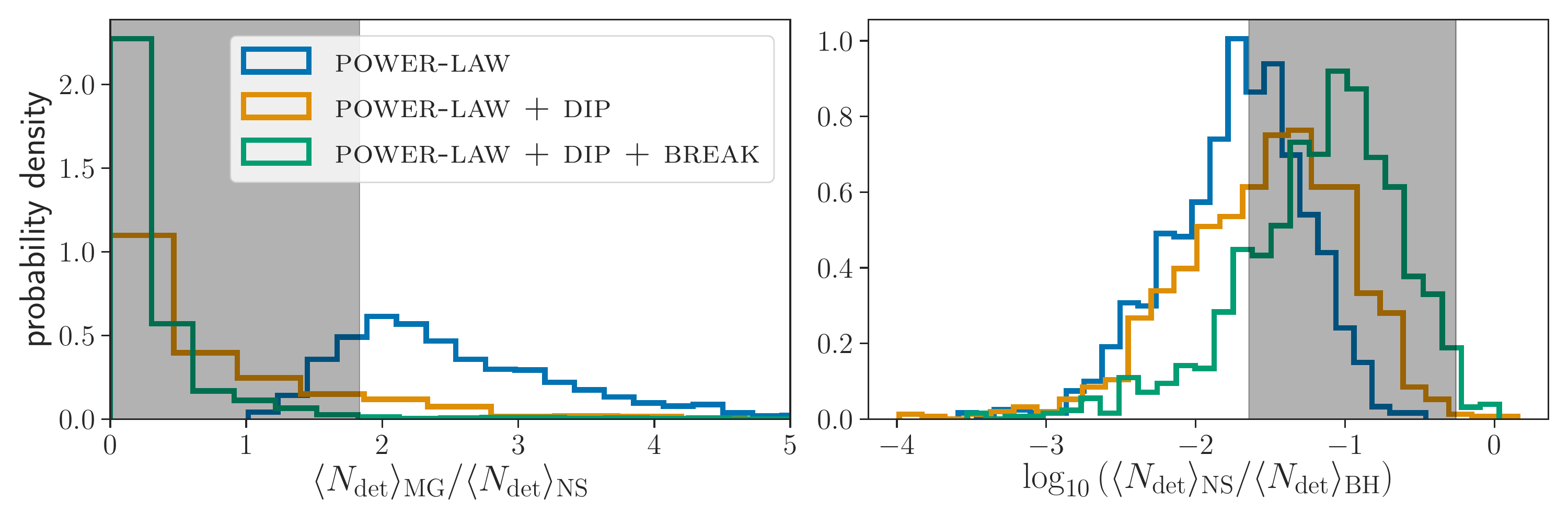}
    \end{center}
    \caption{
        \emph{Left}: Ratio of the expected number of detections with primary mass in the MG (defined here as $2.5$--$5 \,M_\odot$) compared to the expected number of neutron star detections (defined here as $1$--$2.5 \,M_\odot$).
        \emph{Right}: Ratio of the expected number of detections with a neutron star primary mass (in the mass range  $1$--$2.5 \,M_\odot$) compared to a BH primary mass (in the mass range $5 \,M_\odot$--$100 \,M_\odot$).
        GWTC-1 contains 0 events with $m_1$ in the MG, 1 BNS, and 10 BBHs, leading us to measure a MG-NS ratio of $\sim 0$ and a NS-BH ratio of $\sim 0.1$; the 90\% highest posterior density credible intervals on these values are shown in the gray bands (see text for more detail).
        The simple \textsc{power-law} model predicts at least as many mass-gap detections as neutron-star detections.
        Meanwhile, when we allow for both a dip and a break in the power-law, we lower the expected fraction of mass-gap detections, and raise the expected fraction of neutron-star detections relative to black hole detections, allowing us to better fit the observed number of detections in each bin.
    }
    \label{fig:MG-NS-BH-ratios}
\end{figure*}

Another way of evaluating the goodness-of-fit of the models is shown in Fig.~\ref{fig:MG-NS-BH-ratios}, which compares the expected fraction of detections in different primary mass bins---the NS mass range $m_1 \in [1, 2.5] \,M_\odot$, the mass-gap (MG) range $m_1 \in [2.5, 5] \,M_\odot$, and the BH mass range $m_1 \in [5, 100] \,M_\odot$---as predicted from the \textsc{power-law} fit (in blue).
These values are found by integrating the PPD {cumulative distribution functions} of Fig.~\ref{fig:ppc} between the specified mass boundaries.
The boundaries used here are chosen only for illustrative purposes, although the maximum mass achievable by NSs is an area of active study~\citep[e.g.,][]{MMaxmodelsel, 2020arXiv200304880L, 2020arXiv200407744E, PhysRevLett.121.161101, GW170817ModelSelection, Margalit_2017, PhysRevD.100.023015, Rezzolla_2018}.
Regardless of the precise boundaries, the true mass distribution should be able to accurately predict the fraction of events detected in each category.
For the remainder of this work, we will use the NS, MG, and BH labels to refer to these bins in primary mass, unless stated otherwise.

Despite the large measurement uncertainties on primary mass, each of the GWTC-1 events falls clearly into one of these mass-based categories, so that we can trivially count 1 NS, zero MGs, and 10 BH.
Letting $f_\mathrm{NS}$, $f_\mathrm{MG}$ and $f_\mathrm{BH}$ denote the \emph{expected}\/ fraction of detections in the NS, MG, and BH mass range according to the true underlying mass distribution, the \emph{observed}\/ number of detections in each category, out of $N$ total detections, follows a trinomial distribution:
\begin{equation}
    \begin{split}
        p(N_\mathrm{NS}, N_\mathrm{MG}, N_\mathrm{BH} & \mid f_\mathrm{NS}, f_\mathrm{MG}, f_\mathrm{BH}) = \\ & \frac{N!}{N_\mathrm{NS}! N_\mathrm{MG}! N_\mathrm{BH}!} f_\mathrm{NS}^{N_\mathrm{NS}} f_\mathrm{MG}^{N_\mathrm{MG}} f_\mathrm{BH}^{N_\mathrm{BH}}.
    \end{split}
\end{equation}
Given $N_\mathrm{NS} = 1$, $N_\mathrm{MG} = 0$, $N_\mathrm{BH} = 10$, we can calculate the posterior on $f_\mathrm{NS}$, $f_\mathrm{MG}$, and $f_\mathrm{BH}$ according to:
\begin{multline}
    p(f_\mathrm{NS}, f_\mathrm{MG}, f_\mathrm{BH} \mid N_\mathrm{NS}, N_\mathrm{MG}, N_\mathrm{BH}) \propto \\ p(N_\mathrm{NS}, N_\mathrm{MG}, N_\mathrm{BH} \mid f_\mathrm{NS}, f_\mathrm{MG}, f_\mathrm{BH}) \\ \times p_0(f_\mathrm{NS}, f_\mathrm{MG}, f_\mathrm{BH}) .
\end{multline}
We take the Jeffreys prior for $p_0(f_\mathrm{NS}, f_\mathrm{MG}, f_\mathrm{BH})$, which is a symmetric Dirichlet distribution with concentration parameter $\alpha = 0.5$.
The posterior is then given by a Dirichlet distribution with parameters $\mathbf{\alpha} = (N_\mathrm{NS} + 0.5, N_\mathrm{MG} + 0.5, N_\mathrm{BH} + 0.5)$.
The posterior on $f_\mathrm{NS}$, $f_\mathrm{MG}$, and $f_\mathrm{BH}$, produced by drawing from a Dirichlet distribution in this way, is used to produce the gray bands in Fig.~\ref{fig:MG-NS-BH-ratios}.
\added{Here, the \emph{expected number} of events in category $\mathcal{C}$, $\langle N_\mathrm{det} \rangle_\mathcal{C}$, is related to the expected fraction $f_\mathcal{C}$ by $\langle N_\mathrm{det} \rangle_\mathcal{C} = N f_\mathcal{C}$, so $\langle N_\mathrm{det} \rangle_\mathrm{MG} /  \langle N_\mathrm{det} \rangle_\mathrm{NS} = f_\mathrm{MG}/f_\mathrm{NS}$ and so on.}
%\reed{We further note that ratios of the observed counts should match the ratios of the fractions, which is to say that the expected value of $N_\mathrm{MG}/N_\mathrm{NS}$ over many different realizations of detected systems should be $f_\mathrm{MG}/f_\mathrm{NS}$ and so on.}

According to the \textsc{power-law} fit to all 11 GWTC-1 detections, we should detect one BNS system per \result{$48^{+370}_{-38}$} systems containing a BH.
This is in mild tension with our detection of 1 BNS system per 10 BBH systems, which implies that the detection ratio is \result{$\langle N_\mathrm{det} \rangle_\mathrm{BH}/ \langle N_\mathrm{det} \rangle_\mathrm{NS}  =  8.8^{+51.0}_{-6.5}$} (median and 90\% symmetric interval).
Meanwhile, according to the \textsc{power-law} fit, we expect \result{$2.37^{+2.06}_{-0.98}$} systems with a MG primary mass per BNS event.
Again, this is in mild tension with our observation of no systems in the mass-gap and one BNS system, which suggests that the detection ratio \result{$\langle N_\mathrm{det} \rangle_\mathrm{MG}/ \langle N_\mathrm{det} \rangle_\mathrm{NS} < 1.9$} at 90\% credibility.

In summary, we find that within statistical uncertainties, the \textsc{power-law} model provides a marginally adequate fit to the 11 GWTC-1 detections. {However, tensions emerge, hinting at possible features in the mass spectrum between the NS and BH mass range:}
\begin{enumerate}[label=(\alph*)]
    \item {\em GW170817 is a low-mass outlier with respect to the BBH population}. Based on the BBH-only \textsc{power-law} fit, we would expect to detect a system with $m_1 < 2 M_\odot$, given 11 total detections, only \result{0.13\%} of the time. When we update the \textsc{power-law} fit with all 11 detections, the hyper-parameters shift to accommodate GW170817 (see Fig.~\ref{fig:PL_corner_compare}) and this increases to \result{17\%} {of the time}.
    \item {\em The mass-gap is too empty.} The \textsc{power-law} fit to the 11 detections \emph{over}-predicts the number of mass-gap detections compared to NS detections. We would expect to detect a greater number of BNS systems than MG systems, given 11 detections, only \result{12\%} of the time.
    \item {\em GW170817 is a surprise.} The \textsc{power-law} fit \emph{under}-predicts the number of NS detections compared to BH detections: out of 11 total detections, we would expect to detect one NS primary mass and 10 BH primary masses only \result{9\%} of the time based on this fit.
\end{enumerate}
In the following section we characterize possible features between the NS and BH mass spectrum, including a mass gap and a power-law break, and explore how their presence alleviates these tensions.

%-------------------------------------------------

\section{Characterizing a feature between neutron stars and black holes}
\label{sec:results}

The previous section examined the ability of a single power-law to fit the BNS and BBHs of GWTC-1.
Below, we fit the full mass distribution of Eq.~\ref{eq:pm-dipbreak} to the GWTC-1 events.
We investigate the presence of a feature between the NS and BH mass spectrum, quantifying the evidence in favor of a mass dip, gap, or break between NS and BH masses.

The first extension we consider to a power-law mass spectrum is \textsc{power-law + dip}, parametrized by the notch filter of Eq.~\ref{eq:notch}, which suppresses the merger rate for masses $\gamma_\mathrm{low} < m < \gamma_\mathrm{high}$.
The free parameters of this model are the minimum NS mass $\mmin$, maximum BH mass $\mmax$, power-law slope $\alpha$, amplitude of the dip $A$, dip boundaries $\gamma_\mathrm{low}$ and $\gamma_\mathrm{high}$, and mass-ratio power-law slope $\beta$.
While it is important to remember that the lower edge of the dip may or may not correspond to the maximum NS mass, depending on whether BHs exist below the gap, this subtlety does not affect our analysis.
For convenience, we introduce a parameter describing the gap width $w = \gamma_\mathrm{high} - \gamma_\mathrm{low}$, and set flat priors on \result{$\mmin \in [1 \,M_\odot, 1.4 \,M_\odot]$, $\mmax \in [35 \,M_\odot, 100 \,M_\odot]$, $\alpha \in [-5, 2]$, $A \in [0 , 1]$, $\gamma_\mathrm{low} \in [1.4 \,M_\odot, 3 \,M_\odot]$, $w \in [2 \,M_\odot, 6 \,M_\odot]$, and $\beta \in [-4, 12]$}.

With this choice of priors, we allow for a mass gap starting at $1.4 \,M_\odot < \gamma_\mathrm{low} < 3 \,M_\odot$ with a width of $2 \,M_\odot < w < 6 \,M_\odot$.
Our priors on the dip location are externally motivated by observational and theoretical expectations for NS masses \citep[see, e.g..][]{2020arXiv200304880L, 2020arXiv200407744E, 2020RNAAS...4...65F, Ozel}.
We verify that our results are not driven by the prior choice with a ``look-elsewhere'' test, fitting the full \textsc{power-law + dip + break} model to {only} the 10 GWTC-1 BBH detections and finding that, although we cannot rule out the presence of a second dip in the mass spectrum, there is no compelling evidence for a dip in the BBH mass range; we simply recover the prior on $A$.

The fit to the primary mass distribution under \textsc{power-law + dip} is shown in the second panel of Fig.~\ref{fig:dragons}.
Comparing to the top panel, which shows the \textsc{power-law} fit, we can see the data prefer some decrease in the merger rate between \result{$\gamma_\mathrm{low} = 2.2^{+0.6}_{-0.5} \,M_\odot$ and $\gamma_\mathrm{high} = 6.7^{+1.0}_{-1.5}$}.\footnote{The posterior on $\gamma_\mathrm{low}$ is not well-constrained, and follows the prior, which is to be expected from only one detection between $\mmin$ and $\gamma_\mathrm{low}$, and the low sensitivity at low masses.}
The posterior on the amplitude of the dip, $A$, peaks at $A = 1$, corresponding to a {perfect} mass gap, with a tail down to $A = 0$, corresponding to an uninterrupted power-law.
We find that a {perfect} mass gap is preferred over \textsc{power-law} by a factor of \SDDRgap.
\added{As an additional test, we fix $A = 1$ within our parameterization of Eq.~\ref{eq:notch} and allow the sharpness of the gap edges, $\eta_\mathrm{low}$ and $\eta_\mathrm{high}$, to vary between $\eta_\mathrm{low} = \eta_\mathrm{high} = 0$ (no gap) and $\eta_\mathrm{low} = \eta_\mathrm{high} = 50$ (perfect mass gap).\footnote{Recall that our default choice  throughout this work is to fix $\eta_\mathrm{low} = \eta_\mathrm{high} = 50$.} We find that within this model, a perfect gap ($\eta_\mathrm{low} = \eta_\mathrm{high} = 50$) is preferred over an uninterrupted power-law by a Bayes factor of $\result{6.0}$, similar to the preference we recover in the default model.}
\added{
While this is suggestive, noise fluctuations are expected to occasionally produce Bayes factors at least this large (see{\it, e.g.}, \citealt{2014PhRvD..89h2001A}).
Indeed, \citet{Jeffreys1961} suggests that the ratio would need to be $\gtrsim 100$ to be decisive.
}

The second feature we allow in the mass distribution is a break in the power-law in addition to a dip.
The fit to the primary mass distribution under the \textsc{Power-law + dip + break} model is shown in the bottom panel of Fig.~\ref{fig:dragons}.
In this model, we adopt the same priors on the free parameters $\mmin$, $\mmax$, $A$, $\gamma_\mathrm{low}$, $w = \gamma_\mathrm{high} - \gamma_\mathrm{low}$, and $\beta$.
Within the full model, we recover similar constraints on the parameters describing the dip, still favoring a mass gap $A = 1$ over a simple power-law $A = 0$ by a factor of \result{$\sim 4$}.
\added{Again, this is suggestive but not definitive.}

\begin{figure}
    \begin{center}
        \includegraphics[width=1.0\columnwidth, clip=True, trim=10 15 15 00]{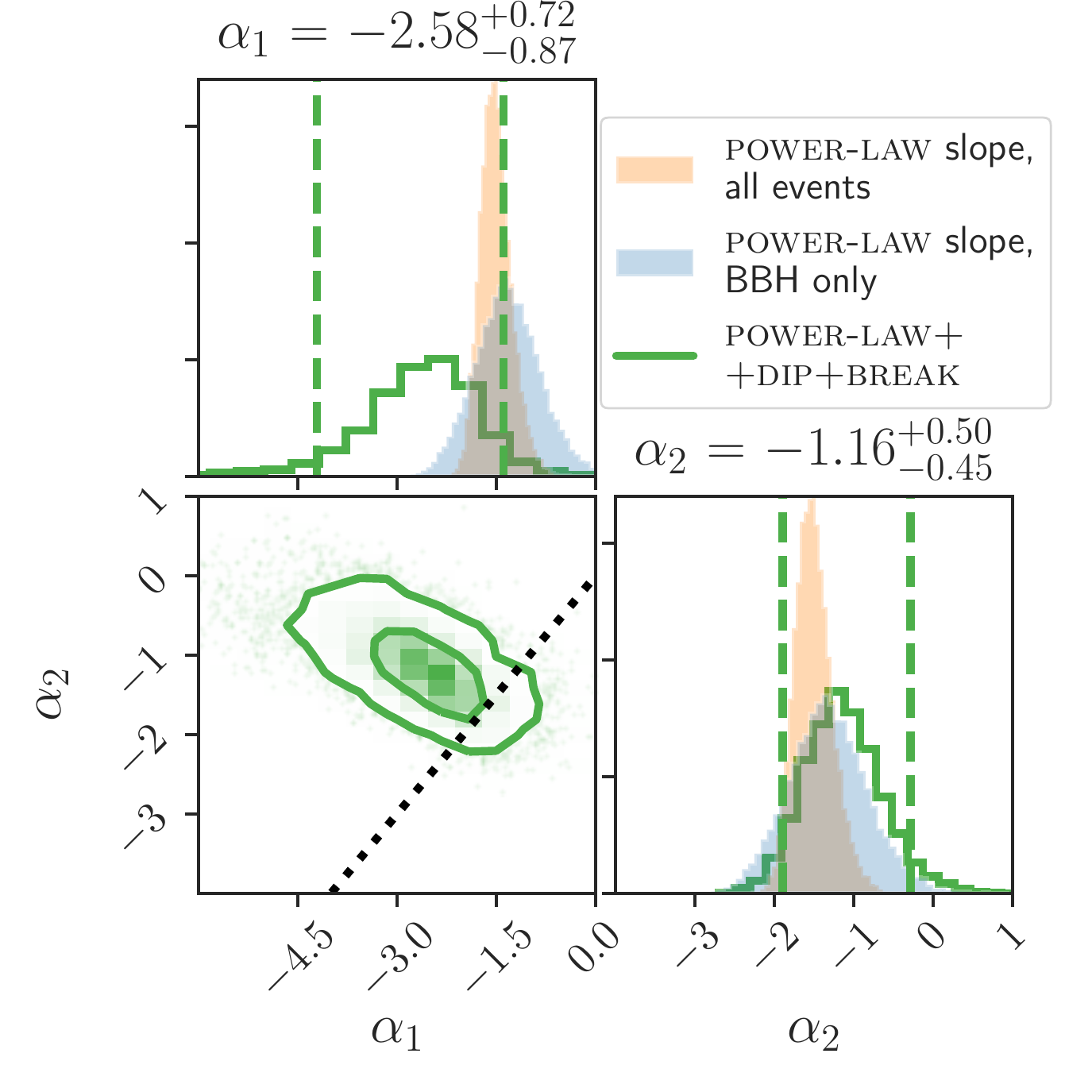}
    \end{center}
    \caption{
        (\emph{green}) The joint posterior on the power-law slopes $\alpha_1$ (NS mass range) and $\alpha_2$ (BH mass range) for the full \textsc{power-law + dip + break} model.
        The two-dimensional contours show 50\% and 90\% credible regions, while vertical dashed lines show one-dimensional 90\% credible intervals.
        For comparison, we show the power-law slope inferred under the \textsc{power-law} fit to {(\emph{blue}) only} the BBH and {to} (\emph{orange}) all GWTC-1 events.
        We recover $\alpha_1 < \alpha_2$ with \Palphagtralpha~credibility; $\alpha_1 = \alpha_2$ in this model reduces to the \textsc{power-law + dip} model.
    }
    \label{fig:alpha-corner}
\end{figure}

The \textsc{Power-law + dip + break} model additionally allows for a break in the power-law at $\gamma_\mathrm{high}$, with power-law slope $\alpha_1$ at $m < \gamma_\mathrm{high}$ and slope $\alpha_2$ at $m > \gamma_\mathrm{high}$.
This allows for the possibility that the NS and BH mass spectra are described by different power-laws, with a possible gap between them.
We set flat, uninformative priors on $\alpha_1$ and $\alpha_2$: \result{$\alpha_1, \alpha_2 \in [-8, 2]$}.
The joint posterior on $\alpha_1$ and $\alpha_2$ can be seen in Fig.~\ref{fig:alpha-corner}.
We find that $\alpha_1$, the NS power-law slope, is likely steeper than $\alpha_2$, the BH power-law slope.
The data prefer $\alpha_1 < \alpha_2$ at \Palphagtralpha~credibility, with \result{$\alpha_1 = -2.58^{+0.72}_{-0.87}$} and \result{$\alpha_2 = -1.16^{+0.50}_{-0.45}$}.
The inferred value of the BH power-law slope, $\alpha_2$, is very similar to the power-law slope inferred with the BBH-only fit, $\alpha = \alphaBHPL$, and prefers to be slightly shallower than the power-law slope inferred under the \textsc{power-law} model fit to all events, $\alpha = \alphaonePL$; see the comparison in Fig.~\ref{fig:alpha-corner}.

We can understand the preference for both a dip and a break between NS and BH masses by returning to the posterior predictive checks of Figs.~\ref{fig:ppc} and~\ref{fig:MG-NS-BH-ratios}.
The simple \textsc{power-law} fit under-predicts the fraction of detections in the NS range, while over-predicting the fraction of detections in the MG range, compared to the current observations of one BNS, zero MGs, and 10 BBH.
In the \textsc{power-law} model, the detection rate of MG detections must be at least \result{$1.5$} times as large as the NS detection rate (at 90\% credibility); increasing the fraction of NSs within the \textsc{power-law} model necessitates an increase in the fraction of MG events.

By introducing a dip, we decrease the expected number of MG detections to \result{$< 1.9$} per NS detection (90\% credibility), while slightly increasing the expected {fraction} of NS detections to one NS detection per \result{$28^{+288}_{-22}$} BH detections.

Introducing a break in the power-law in addition to a dip allows us to further increase the expected number of NS detections to 1 NS detection per \result{$13^{+141}_{-10}$} BH detections, bringing it close to the GWTC-1 observation of 1 BNS per 10 BBH detections, while maintaining a low rate of mass-gap detections (\result{$<0.80$} per NS detection).

%-------------------------------------------------
\section{Discussion}
\label{sec:discussion}

Given the evidence for a feature between NSs and BHs, we now consider several {implications, including delineating our knowledge} about specific objects in \S\ref{sec:single-event}, updated  astrophysical rates in \S\ref{sec:rates}, and prospects for the coming years in \S\ref{sec:future}.

%------------------------
\subsection{Updated single-event classification}
\label{sec:single-event}

Although not conclusive, the GWTC-1 detections show hints of a feature between NS and BH masses.
Future detections will allow us to resolve this feature with increased precision, which may provide a natural boundary between the NS and BH populations.
Meanwhile, our inference of the compact object mass distribution allows us to update the mass measurements of individual events, often allowing for much tighter constraints than the posteriors inferred under uninformative priors~\citep{2020ApJ...891L..31F,2019arXiv191209708G,2020arXiv200106051M}.
For example, if the population fit reveals a mass gap between NS and BH masses, applying the population prior to events for which the likelihood measurement uncertainty is broad and overlaps with the gap will significantly tighten the mass posteriors, forcing the posterior support to lie below or above the gap~\citep{2020ApJ...891L..31F}.
Simultaneously fitting the population distribution and the masses of events can self-consistently classify detected sources into {NSs and BHs}~\citep{2015PhRvD..91b3005F,2015MNRAS.450L..85M}.
{However, we note that the feature that emerges in the mass distribution may not necessarily correspond to the boundary between NS and BH masses, but instead may be complicated by accretion, hierarchical mergers, or primordial} black holes~\citep{2016PhRvD..94h3504C,2018ApJ...856..110Y}.
External priors on the NS maximum mass may also be applied, together with the population fit, in order to aid in the classification~\citep{GW190814,MMaxmodelsel}.

%------------------------
\subsection{Compact object merger rates}
\label{sec:rates}

Regardless of whether there exists a mass feature that naturally distinguishes {between} sub-populations of compact objects, our fit to the full mass distribution allows us to derive the compact object merger rate in different mass bins without explicitly counting the number of events in each category~\citep{2015PhRvD..91b3005F,2020CQGra..37d5007K}.
Defining the BNS category as $1 \,M_\odot < m_2 < m_1 < 2.5 \,M_\odot$ and the BBH category as $5 \,M_\odot < m_2 < m_1 < 100 \,M_\odot$, we calculate the merger rate for each category by integrating the inferred rate density {$d{N}/ dV_c dt_s dm_1 dm_2$} under the \textsc{power-law}, \textsc{power-law + dip}, and \textsc{power-law + dip + break} models within the specified $(m_1, m_2)$ region.

\begin{figure}
    \begin{center}
        \begin{tikzpicture}
           \draw (0, 0) node {\includegraphics[width=1.0\columnwidth, clip=True, trim=20 15 20 15]{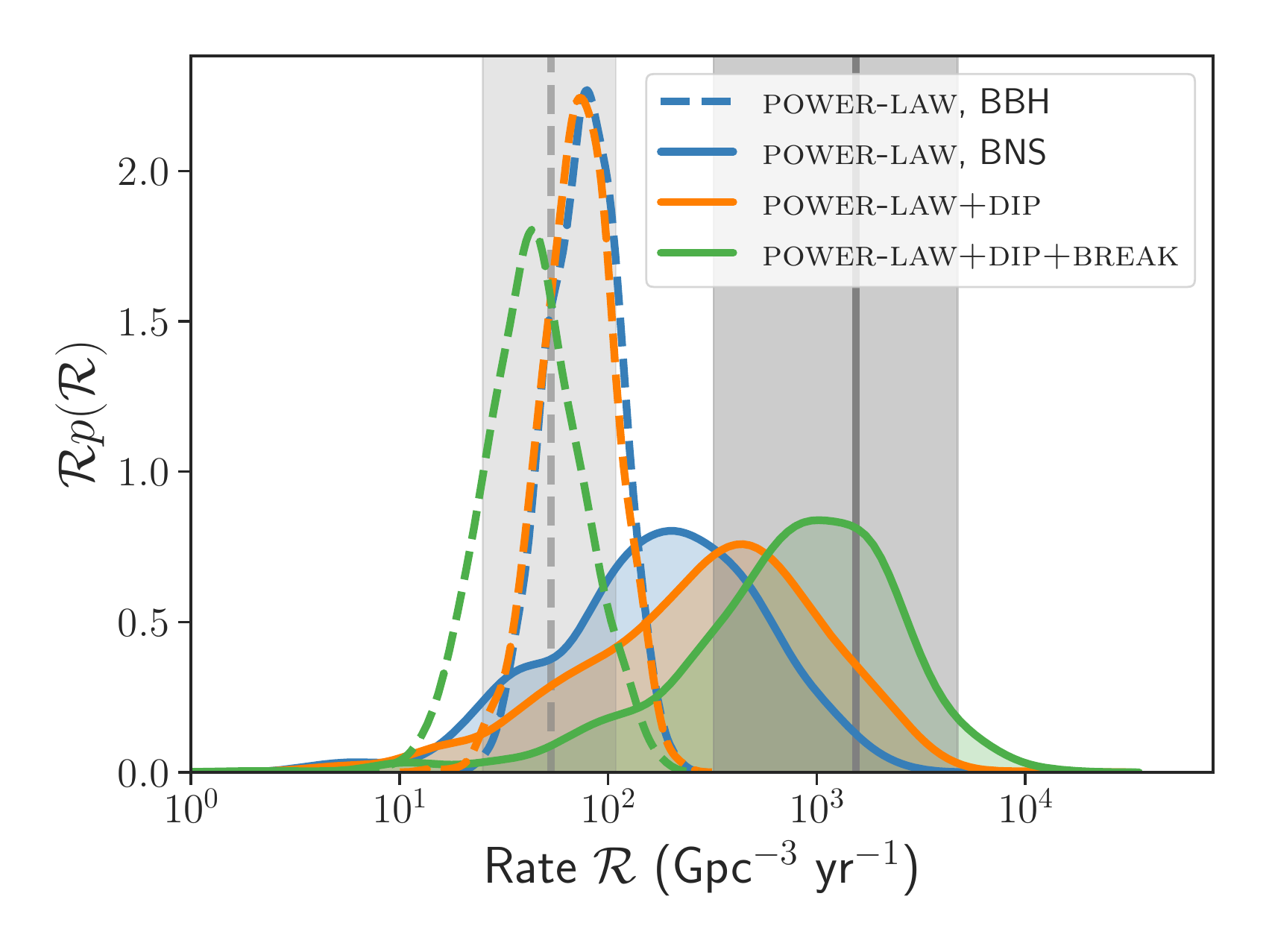}};
           \draw (-1.8, -0.6) node {{\large{BBH}}};
           \draw (+2.4, -0.6) node {{\large{BNS}}};
        \end{tikzpicture}
    \end{center}
    \caption{
        Astrophysical merger rate within two different mass bins: for BNS, $1 \,M_\odot < m_2 < m_1 < 2.5 \,M_\odot$, and for BBH, $5 \,M_\odot < m_2 < m_1 < 100 \,M_\odot$, as inferred by each of the models.
        The dashed, unfilled probability density curves centered below $\mathcal{R}\sim 10^2\, \rateunits$ show the BBH rate inference, while the solid, filled curves show BNS rate inference, for the \textsc{power-law} (blue), \textsc{power-law + dip} (\emph{orange}), and \textsc{power-law + dip + break} (\emph{green}).
        For comparison, the gray bands show the median and 90\% credible intervals of the BBH rate (\emph{dashed}) and BNS rate (\emph{solid}) inferred by the LVC in~\cite{2019PhRvX...9c1040A,2019ApJ...882L..24A}.
        Allowing for a dip and break between NS and BH masses tends to decorrelate the merger rates, increasing the inferred BNS merger rate and decreasing the BBH merger rate.
    }
    \label{fig:rate}
\end{figure}

The results are shown in Fig.~\ref{fig:rate}.
In the \textsc{power-law} model, the BNS and BBH merger rates are closely correlated, while adding the features of the \textsc{power-law + dip} and \textsc{power-law + dip + break} increases the BNS merger rate estimate while decreasing the BBH rate estimate.
The BBH rate under the \textsc{power-law} model is $\mathcal{R}_\mathrm{BBH} = \BBHratePL \rateunits$, while the full \textsc{power-law + dip + break} model yields $\BBHratedipbreak \rateunits$.
Meanwhile, the inferred BNS merger rate is $\mathcal{R}_\mathrm{BNS} = \BNSratePL \rateunits$ under \textsc{power-law} and $\BNSratedipbreak \rateunits$ under \textsc{power-law + dip + break}.
The full \textsc{power-law + dip + break} model better matches the rate estimates of~\cite{2019PhRvX...9c1040A,2019ApJ...882L..24A}, which assumed separate BNS and BBH populations, of $\mathcal{R}_\mathrm{BBH} = \LVCBBHrate \rateunits$ and $\mathcal{R}_\mathrm{BNS} = \LVCBNSrate \rateunits$.
These trends for the astrophysical rates are consistent with the detection rates explored in Fig.~\ref{fig:MG-NS-BH-ratios}.

We can also extrapolate our models to calculate the rate in the NSBH category ($ 5 \,M_\odot < m_1 < 100 \,M_\odot$, $1 \,M_\odot < m_2 < 2.5 \,M_\odot$), although we caution that this is a significant extrapolation since our simple pairing function of Eq.~\ref{eq:pm-2d} may not apply to the NSBH mass region.
Because the GWTC-1 detections are all consistent with having equal component masses, our fits strongly disfavor unequal mass pairings, and predict a low NSBH rate, with an upper 95\% limit of \result{$8.2 \rateunits$}.

We reiterate that our choice of mass bins to classify NS, MG and BH sources, and calculate the corresponding rates, is only illustrative.
Future detections will enable us to set the mass bins according to the measured feature in the mass distribution, or external measurements of the NS maximum mass, while accounting for uncertainty in the bin edges~\citep{MMaxmodelsel}.
Additional detections will also allow us to meaningfully constrain the rate as a function of redshift~\citep{2018ApJ...863L..41F,2019ApJ...882L..24A}.

%------------------------

\subsection{Looking ahead}
\label{sec:future}

As seen in \S\ref{sec:results}, the preference for a dip/break between NS and BH masses can be understood by the dearth of detections between $\sim 2 \,M_\odot$--$8 \,M_\odot$, relative to the number of detections below the purported mass gap (GW170817) and above the gap (10 BBHs) {in} GWTC-1.
Recall that, defining the NS range as $m_1 \in [1, 2.5] \,M_\odot$, the MG range as $m_1 \in [2.5, 5] \,M_\odot$, and the ``low-mass" BH range as $m_1 \in [5, 10] \,M_\odot$, the \textsc{power-law} fit would have us detect \result{$2.40^{+1.95}_{-1.02}$} MG systems  for every BNS system, and 1 MG system for every \result{$2.33^{+1.92}_{-0.91}$} low-mass BH system (90\% credibility).
For a wider MG region defined between $2.5$--$7 \,M_\odot$, we expect \result{$4.63^{+5.55}_{-2.33}$} MG systems per BNS system, and \result{$1.39^{+0.79}_{-0.54}$} MG systems per low-mass BH systems ($7 \,M_\odot < m_1 < 10 \,M_\odot$).

While the current preference for a dip/break is suggestive, with only 11 events the statistical uncertainties remain large.
The situation will improve with future detections.
For example, with 100 detections the featureless \textsc{power-law} model would require at least \result{$N_\mathrm{BNS}-2$} detections with true primary masses in the mass range $2.5$--$5 \,M_\odot$, where $N_\mathrm{BNS}$ is the number of detections in the range $1$--$2.5 \,M_\odot$ (95\% credibility).
For a wider MG region between $2.5$--$7 \,M_\odot$, \textsc{power-law} predicts \result{at least $N_\mathrm{BNS} + 1$ MG detections}.
Additionally, the \textsc{power-law} model predicts no more than \result{$9$ BNS detections given a total of 100 detections}.
Observing a smaller number of MG detections or a larger number of BNS detections would provide {further} evidence for a feature between the NS and BH mass range.

Of course, it is important to carry out the full population analysis that takes into account the measurement uncertainties of detected systems.
Even in the presence of an absolute gap between NS and BH, some \emph{observed} masses will be scattered into the gap due to noise fluctuations~\citep{2017MNRAS.465.3254M,2020ApJ...891L..31F}.
For the power-law type of distributions explored here, we expect \result{$\sim 2\%$ of detected BHs to lie within $1\sigma$ of the mass gap} and \result{$\sim 45\%$ of detected NSs to lie within $1\sigma$ of the gap}, assuming $1\sigma$ measurement uncertainties of $20\%$.
For a gap width $w > \sigma \gtrsim 1 \,M_\odot$, we expect that \result{$\sim 50\%$ of the masses close to the gap on either side will be erroneously observed within the gap}, or \result{$\sim 1\%$ of BH and $\sim 20\%$ of the detected NS primary masses}.
If there exists an absolute mass gap and we observe five detections with NS primary masses {along with} 50 detections with BH primary masses, we would expect \result{$\sim 1.5$ erroneous observations in the gap}, while the expectation from a continuous power-law would be \result{$\sim 10 \pm \sqrt{10}$ detections}, making it possible to identify the presence of a mass gap at \result{$\sim 3\sigma$} with fewer than 100 detections (see a similar argument in~\citealp{2015MNRAS.450L..85M}). These predictions are consistent with the population analysis of~\citet{2017MNRAS.465.3254M}, which employed a clustering analysis on simulated data and found that $\sim 20$ detections on either side of the mass gap would enable its confident detection.

In summary, the ability of future detections to precisely measure features in the mass spectrum depends on the depth and width of the purported mass gap, as well as the sharpness of the features relative to the uncertainty of the observed masses.
If the features are sharp, we expect to converge on their location relatively quickly, scaling with the number of detections $N$ as $N^{-1}$, but if they are less abrupt, we expect to converge as $N^{-0.5}$~\citep{2003Natur.424...42C,2015MNRAS.450L..85M}.

The discussion throughout this paper has focused mainly on the one-dimensional primary mass distribution, because the GWTC-1 events all consist of nearly equal component masses.
In the future, looking for structure in the two-dimensional mass distribution will be important for characterizing a potential population of NSBH systems.
The approach described here to simultaneously fit the mass distribution of all compact objects in binary systems will allow us to explore how the component masses of NSBH systems relate to the component NSs and BHs found in BNS and BBH systems.

%-------------------------------------------------

\section{Conclusion}
\label{sec:conclusion}

This work presents the first analysis to jointly fit the mass distribution of NSs and BHs in merging binary systems using data from the first two observing runs of Advanced LIGO/Virgo.
We assume no external knowledge of NS and BH sub-populations, and ask whether GW170817, the least massive event detected, can be identified as an outlier in the {BBH-only} population based only on its mass, the property that is easiest to measure {with} GWs.
We find that in the context of the BBH population, the masses of GW170817 are exceptional; based on the BBH-only fit, we would expect to detect an event with $m_1 < 2 \,M_\odot$ out of a set of 11 events only \result{0.13\%} of the time.

We next try to fit a continuous power-law across the entire mass range, finding that it is \emph{possible} to {extend} the {\textsc{power-law}} fit to the BBH population of~\cite{2019ApJ...882L..24A} down to the masses of GW170817, but some tensions emerge. Namely, the power-law fit under-predicts the number of detections in the BNS mass range, while over-predicting the number of detections in the MG range.

While more events are required to judge whether the tensions in the {\textsc{power-law}} fit are statistically significant, we find that these tensions can be alleviated by allowing for a dip and/or a break in the mass distribution between NS and BH masses. When we include the possibility of a dip, we find that a mass gap of width $ > 2 \,M_\odot$ is preferred over an uninterrupted power-law by a factor of $\SDDRgap{}$ in GWTC-1. When we further allow the power-law to take a different slope $\alpha_1$ at low (NS) masses, compared to $\alpha_2$ at high (BH) masses, we find that this is also preferred, with $\alpha_1$ steeper than $\alpha_2$ at \result{91\%} credibility.

Considering only the 11 {GW} events from O1+O2, we find preliminary evidence for two distinct populations of sources, with hints of a gap in between.
Jointly fitting the mass distribution of NSs and BHs in binary systems allows us to self-consistently calculate merger rates in different categories and pool our knowledge regarding NS and BH masses across BNS, BBH, and NSBH systems.
\added{The methods described here will provide important insights going forward, especially in light of the latest discoveries from LIGO/Virgo. These include a high-mass BNS system with a total mass of $\sim 3 \ M_\odot$~\citep{GW190425} and a highly asymmetric BBH or NSBH system with a secondary mass of $\sim 2.6 \ M_\odot$~\citep{GW190814}.}
With 100 events, as might be expected by the end of O3 or early in O4 \citep{2018LRR....21....3A}, GW data alone may provide a clear indication of the existence of separate NS and BH populations, as well as provide important constraints on the existence and associated parameters of a mass gap between NSs and BHs.
%-------------------------------------------------

\acknowledgments
We are grateful to Katerina Chatziioannou, Will Farr, and Chase Kimball for their helpful comments on the manuscript.
MF was supported by the NSF Graduate Research Fellowship Program under grant DGE-1746045.
MF and DEH were supported by NSF grant PHY-1708081.
MF, RE, and DEH were supported by the Kavli Institute for Cosmological Physics at the University of Chicago and an endowment from the Kavli Foundation and its founder Fred Kavli.
DEH gratefully acknowledges a Marion and Stuart Rice Award.
The authors also acknowledge the computational resources provided by the LIGO Laboratory and supported by NSF grants PHY-0757058 and PHY-0823459.
This research has made use of data, software and/or web tools obtained from the Gravitational Wave Open Science Center (https://www.gw-openscience.org), a service of LIGO Laboratory, the LIGO Scientific Collaboration and the Virgo Collaboration. LIGO is funded by the U.S. National Science Foundation. Virgo is funded by the French Centre National de Recherche Scientifique (CNRS), the Italian Istituto Nazionale della Fisica Nucleare (INFN) and the Dutch Nikhef, with contributions by Polish and Hungarian institutes.

%-------------------------------------------------
\bibliographystyle{aasjournal}
\bibliography{references}

%-------------------------------------------------

\appendix
\section{Analysis details}
\label{sec:statmethods-appendix}
We write the number density of sources as:
\begin{equation}
\label{eq:numberdensity}
\frac{d{N}}{dz dm_1 dm_2} = {\mathcal{N}} p(z) p(m_1, m_2 \mid \Lambda),
\end{equation}
where the merger rate density is given by:
\begin{equation}
\left. \frac{d{N}}{dm_1 dm_2 dV_c dt_s} \right|_z = \frac{d{N}}{dm_1 dm_2 dz} \times (T_\mathrm{obs}/(1+z))^{-1} \times (dV_c/dz)^{-1},
\end{equation}
with $V_c$ the comoving volume, $t_s$ the time as measured in the source-frame, and $T_\mathrm{obs}$ the total observing time, or 169.7 days for O1+O2~\citep{2019PhRvX...9c1040A}.
We assume that the rate density $\mathcal{R}$ is constant in redshift, consistent with the GWTC-1 detections~\citep{2018ApJ...863L..41F,2019ApJ...882L..24A}, so that the normalization term $\mathcal{N}$ is related to the astrophysical merger rate $\mathcal{R}$ by:
\begin{equation}
\label{eq:NtoR}
\mathcal{N}= T_\mathrm{obs}\mathcal{R} \int_0^{z_\mathrm{max}}dz(dV_c/dz) (1+z)^{-1},
\end{equation}
and the redshift distribution is:
\begin{equation}
p(z) = \frac{(dV_c/dz) (1+z)^{-1}}{\int_0^{z_\mathrm{max}}dz(dV_c/dz) (1+z)^{-1}}.
\end{equation}
For convenience we denote $A(z_\mathrm{max}) = \int_0^{z_\mathrm{max}}dz(dV_c/dz) (1+z)^{-1}$, and we pick $z_\mathrm{max} = 1$, as no sources are detectable beyond this redshift for O1/O2~\citep{2018LRR....21....3A}.
Marginalizing away the normalization term $\mathcal{N}$ with a prior $p(\mathcal{N})\propto \mathcal{N}^{-1}$, we obtain a posterior for the population hyper-parameters ($\Lambda$) given the observed data $\{\mathcal{D}_i\}$~\citep{2019MNRAS.486.1086M}:
\begin{equation}
\label{eq:popposterior}
    p(\Lambda|\{\mathcal{D}_i\}) = p(\Lambda) \prod\limits_i^N \frac{Z(\mathcal{D}_i|\Lambda)}{\beta(\Lambda)},
\end{equation}
where:
\begin{equation}
    Z(\mathcal{D}_i|\Lambda) \equiv \frac{1}{A(z_\mathrm{max})} \int dm_1 dm_2 dz\, \left[ p(m_1, m_2|\Lambda) \times \left(\frac{1}{1+z}\right)\left(\frac{dV_c}{dz}\right)
p(\mathcal{D}_i|m_1, m_2, z) \right]
\end{equation}
is the marginal likelihood for the $i^\mathrm{th}$ event, and
\begin{equation}
\label{eq:beta}
    \beta(\Lambda) \equiv \frac{1}{A(z_\mathrm{max})} \int dm_1 dm_2 dz\, \left[ p(m_1, m_2|\Lambda)
    \times\left(\frac{1}{1+z}\right) \left(\frac{dV_c}{dz}\right) P(\mathrm{det}|m_1, m_2, z) \right]
\end{equation}
is the fraction of events we expect to detect in a population described by $\Lambda$.
The term $P(\mathrm{det}|m_1, m_2, z)$ in Eq.~\ref{eq:beta} is the probability of detecting a specific system with component masses $m_1$ and $m_2$ at redshift $z$. We calculate this term following the semi-analytic calculation described in~\citet{2016PhRvX...6d1015A,2019ApJ...882L..24A}, which assumes that sources are detected if they have a single-detector signal-to-noise ratio (SNR) $\rho > 8$, calculated with the \emph{Advanced LIGO Early High Sensitivity}  noise power spectral density~\citep{2018LRR....21....3A}.

Note that in addition to the fixed redshift distribution, we assume a fixed spin distribution (uniform spin magnitudes with isotropic orientations) and focus on only the mass distribution.
As such, we neglect the possible impact of spins within the selection function $P(\mathrm{det}|m_1, m_2, z)$, as detectability is predominantly determined by the component masses and the redshift.

Sampling from the posterior distribution of Eq.~\ref{eq:popposterior} produces our main results regarding the \emph{shape} of the mass distribution.
However, we can obtain estimates of the overall rate from:
\begin{align}
    p(\mathcal{N}|\{\mathcal{D}\}) & = \int d\Lambda\, p(\mathcal{N}|\Lambda, \{\mathcal{D}_i\}) p(\Lambda|\{\mathcal{D}_i\}) \nonumber \\
                                   & = \int d\Lambda\, \frac{1}{\mathcal{N}} \mathcal{N}^{N_\mathrm{obs}} e^{-\mathcal{N} \beta(\Lambda)} p(\Lambda|\{\mathcal{D}_i\}) \label{eq:rates}
\end{align}
and use Eq.~\ref{eq:NtoR} to convert $\mathcal{N}$ to the astrophysical merger rate $\mathcal{R}$.
Here, $N_\mathrm{obs}$ is the number of observations, and we have again assumed a prior $p(\mathcal{N}) \propto \mathcal{N}^{-1}$.

We estimate $Z(\mathcal{D}_i|\Lambda)$ by reweighting publicly available posterior samples~\citep{2019arXiv191211716T} that were originally drawn assuming a prior $p_0(m_1, m_2, z)$:
\begin{equation}
    Z(\mathcal{D}_i|\Lambda) \propto \frac{1}{N_i} \sum\limits_j^{N_i} \frac{p(m_1^{(j)}, m_2^{(j)}|\Lambda) (1+z^{(j)})^{-1} (dV_c/dz|_{z^{(j)}})}{p_0(m_1^{(j)}, m_2^{(j)}, z^{(j)})}
\end{equation}
where $m_1^{(j)}, m_2^{(j)}, z^{(j)} \sim p(\mathcal{D}_i|m_1, m_2, z) p_0(m_1, m_2, z)$.
For the samples of GWTC-1, the single-event sampling prior is~\citep{2019ApJ...882L..24A}:
\begin{equation}
    p_0(m_1, m_2, z) \propto d_L(z)^2(1+z)^2\left(d_C(z)+\frac{(1+z)d_H}{E(z)}\right),
\end{equation}
where $d_C$ is the comoving distance, $d_H = c/H_0$ is the Hubble distance, and $E(z) = H(z)/H_0$~\citep{Hogg:cosmo}.

When carrying out posterior predictive checks, we estimate the EDF by drawing samples from the single-event posterior.
For each draw $\Lambda$ of the hyper-posterior, we draw a sample from each of the 11 single-event posteriors, with weights $p(m_1, m_2 \mid \Lambda)p(z) / p_0(m_1, m_2, z)$~\citep{2020ApJ...891L..31F,2019arXiv191209708G, 2020arXiv200106051M}.

Meanwhile, we calculate $\beta(\Lambda)$ with a Monte-Carlo integral over a population of $N_\mathrm{inj} = 2^{26}$ simulated signals $m_1^{(j)}, m_2^{(j)}, z^{(j)} \sim p_\mathrm{draw}(m_1, m_2, z)$~\citep{2018CQGra..35n5009T,Farr_2019}:
\begin{equation}
    \beta(\Lambda) = \frac{1}{N_\mathrm{inj}}\frac{1}{A(z_\mathrm{max})} \left. \sum\limits_k^{N_\mathrm{inj}} \right[ P(\mathrm{det}|m_1^{(k)}, m_2^{(k)}, z^{(k)})
         \left. \times \frac{p(m_1^{(k)}, m_2^{(k)}|\Lambda) (1+z^{(k)})^{-1} (dV_c/dz|_{z^{(k)}})}{p_\mathrm{draw}(m_1^{(k)}, m_2^{(k)}, z^{(k)})} \right],
\end{equation}
accounting for the uncertainty in the Monte-Carlo integral according to \citet{Farr_2019}.
To sample efficiently, $p_\mathrm{draw}$ should resemble the true population distribution.
We pick:
\begin{equation}
p_\mathrm{draw}(m_1, m_2, z) = \frac{1}{A(z_\mathrm{max})}(1+z)^{-1}(dV_c/dz) p_\mathrm{draw}(m_1, m_2),
\end{equation}
with:
\begin{equation}
p_\mathrm{draw}(m_1, m_2)  = \sum_{i = 1}^{2} {f_i} p_\mathrm{PL}(m_1 \mid \alpha_i, \mmin = 1, \mmax^i)
 p_\mathrm{PL}(m_2 \mid \alpha = 0, \mmin = 1, \mmax = m_1),
\end{equation}
with $f_i = [0.5, 0.5]$, $\alpha_i = [-4,-2]$, and $\mmax^i = [10, 100]$.
We calibrate $\beta(\Lambda)$ to the actual selection function from the O1/O2 search~\citep{2016CQGra..33u5004U,2019arXiv190108580S} by dividing the above calculation of $\beta(\Lambda)$ by a constant factor of 1.7; see Fig. 9 in~\citet{2019ApJ...882L..24A}.
\citet{2020ApJ...891L..27F} show that this constant calibration factor is sufficient in recovering the population results of \citet{2019ApJ...882L..24A}.
This same set of injections is used to estimate PPDs, which are found by reweighting the injections according to $p(m_1, m_2, \Lambda)/p_\mathrm{draw}(m_1, m_2)$ with $\Lambda$ drawn from the hyper-posterior of a given model.
We sample from the population hyper-posterior using PyMC3~\citep{Salvatier2016}.

\end{document}